# Three-Leaf Dart-Shaped Single-Crystal BN Formation Promoted by Surface Oxygen

Hui Yang[1,†], Jin Yang[2,†], Xibiao Ren[3,†], Haiyuan Chen[2], Chennupati Jagadish[5], Guang-Can Guo[1,4], Chuanhong Jin[3,a)], Xiaobin Niu[2,b)], and Guo-Ping Guo[1,c)]

[1] Laboratory of Quantum Information, University of Science and Technology of China, Chinese Academy of Sciences, JinZhai Road 96, Hefei 230026, China.
[2] School of Materials and Energy, University of Electronic Science and Technology of China, No. 4, Section 2, North Jianshe Road, Chengdu 610054, China.
[3] State Key Laboratory of Silicon Materials, School of Materials Science and Engineering, Zhejiang University, Zhe Da Road 38, Hangzhou 310027, China.
[4] Institute of Fundamental and Frontier Science, University of Electronic Science and Technology of China, No. 4, Section 2, North Jianshe Road, Chengdu 610054, China.
[5] Department of Electronic Materials Engineering, Research School of Physics and Engineering, The Australian National University, Canberra, ACT 2601, Australia
† These authors contributed equally to this work.
**Electronic mail:** a)gpguo@ustc.edu.cn; b)xbniu@uestc.edu.cn; c)chhjin@zju.edu.cn.

Two-dimensional hexagonal boron nitride (h-BN) single crystals with various shapes have been synthesized by chemical vapor deposition over the past several years. Here we report the formation of three-leaf dart (3LD)-shaped single crystals of h-BN on Cu foil by atmospheric-pressure chemical vapor deposition. The leaves of the 3LD-shaped h-BN are as long as 18 μm and their edges are smooth armchair on one side and stepped armchair on the other. Careful analysis revealed that surface oxygen plays an important role in the formation of the 3LD shape. Oxygen suppressed h-BN nucleation by passivating Cu surface active sites and lowered the edge attachment energy, which caused the growth kinetics to change to a diffusion-controlled mode.

## Introduction

Two-dimensional (2D) materials such as graphene[1-5] and $MoS_2$[6-8] have attracted substantial research attention over the past decade. In particular, the III-V compound hexagonal boron nitride (h-BN) is known as "white graphene" because of its similar lattice structure to that of graphene.[9] Increasing attention is being paid to h-BN because of its various outstanding properties[10,11] such as low dielectric constant,[12] high mechanical strength,[13] high thermal conductivity,[13] and chemical inertness.[14,15] Furthermore, atomically thin h-BN has a very smooth surface with no dangling bonds or surface trapped charges, which can cause carrier scattering.[16,17] In addition, h-BN is an insulator with a large direct band gap (5.97 eV) and very small lattice mismatch (~2%) with that of graphene, making it an ideal dielectric substrate for high-performance graphene devices.[18,19] The electron mobility of a graphene film on an h-BN substrate is much higher than that of one on an $SiO_2$/Si substrate.[18,19] Because of the favorable properties of h-BN and its suitability for various applications, many researchers have attempted to synthesize high-quality, large single crystals of h-BN with uniform thickness.

Similar to graphene, many methods to synthesize h-BN have been developed, such as exfoliating h-BN powder in solution,[13,20-22] unzipping boron nitride nanotubes,[23] and chemical vapor deposition (CVD).[24,25] Among these methods, CVD is the most promising to fabricate large-area films and control the nucleation and growth of h-BN, which is very important to achieve large, high-quality single crystals. Considerable effort has been devoted to synthesizing h-BN by atmospheric-pressure chemical vapor deposition (APCVD) and low-pressure CVD on various substrates, such as Pt,[26,27] Ni,[24] Cu,[25,28-31] and Ru.[32] In these studies, parameters including pressure, hydrogen concentration, substrate, and temperature have been investigated and some useful information has been obtained. For example, triangular-shaped[30] crystalline h-BN flakes are normally produced, although other shapes such as diamonds,[29] hexagons,[28] and some unusual shapes[33] have also been obtained under different conditions. Furthermore, Sushant Sonde et al[54] have indicated role of bulk diffusion and segregation from Ni and Co substrate in the thin film growth of h-BN. Nonetheless, the morphology of h-BN tends to differ between laboratories[28-30] with no plausible explanation, suggesting that there are still parameters affecting the growth process that remain uncontrolled or unknown.

In this work, we report a process to synthesize three-leaf dart (3LD)-shaped h-BN on Cu foil using an APCVD technique. The 3LD-shaped domains are shown to be single crystals with two types of armchair edges (one smooth armchair edge and one stepped armchair edge). The center line of the leaves reaches up to 18 μm long, which is 5 to 6 times larger than the longest side of triangular-shaped flakes grown on the same copper substrate. First principles calculations are used to investigate the effect of oxygen on the formation of 3LD-shaped crystals.

## Results and Discussion

The BN samples were synthesized by APCVD using a solid borane-ammonia complex ($BH_3$-$NH_3$) as the precursor. Before BN growth, the Cu foil substrates were pre-treated by different methods to give two kinds of substrates: low-oxygen Cu (LO-Cu) and high-oxygen Cu (HO-Cu) (see the **Supporting Information** for details). After the pre-treatment of the Cu foil, X-ray photoelectron spectroscopy (XPS) was used to measure the surface oxygen contents of the substrates (Fig. S2). Fig. 1a shows the Cu 2p spectra for LO-Cu and HO-Cu substrates. Compared with that of LO-Cu, the surface oxygen content of HO-Cu was much higher. In the XPS spectrum (Fig. 1a) of HO-Cu, "shake-up" peaks with higher binding energy as compared to that of LO-Cu were observed, which unambiguously acted as a feature of the 2p3/2 spectrum for strongly oxidized Cu species (Cu (II)). In the spectrum of LO-Cu, besides the 2p1/2 (~952.5eV) and 2p3/2 (~932.7eV) peaks from pure Cu, a very weak satellite at ~945 eV originated from remaining oxidized Cu was observed. The original chemical state of LO-Cu may be a mixture of Cu(I) oxide and Cu metal, whereas the original chemical state in HO-Cu is Cu(II).

After growth, the h-BN films were transferred onto 300-nm $SiO_2$/Si substrates and then XPS was used to characterize their chemical compositions. The two characteristic XPS peaks at 398.1 and 190.5 eV in Fig. 1b and c can be attributed to N 1s and B 1s, respectively, in good agreement with other reports.[34,35]

The N:B ratio of the films is 1:1.2. Electron energy loss spectroscopy (EELS) was employed to characterize the chemical bonds of h-BN using samples that were transferred onto a TEM grid coated with an ultrathin carbon membrane. The EELS spectrum contained signals from B and N at 191 and 404 eV, respectively (Fig. 1d), which correspond to the characteristic K-shell excitations of B and N.[36,37] The first peak of two bonds shows the 1s-$\pi^*$ antibonding orbital of each element. Based on these EELS edge structures, the B and N in the BN samples are $sp^2$ hybridized.

Next, we used scanning electron microscopy (SEM) to observe the surface morphology of h-BN. The detailed results are shown in Fig. 2. The h-BN flakes grown on the LO-Cu substrate have a regular triangle (RT) shape (Fig. 2a), which is similar to that observed in a previous study.[30] The BN grown on the HO-Cu substrate (Fig. 2b) consists mostly of flakes with an unusual shape, which we call 3LD, as shown in Fig. 2c. These 3LD-shaped h-BN flakes have threefold rotation symmetry, with angles between two branch center lines of nearly 120°. The branches of most 3LD-shaped h-BN crystals were up to 15 μm long, which is 5 to 6 times longer than the sides of the RT flakes grown on the HO-Cu substrate under the same growth conditions. Besides these 3LD-shaped flakes, we also found some other flakes with needle and V shapes, which can be regarded as components of the 3LD, as shown in Fig. 2d.

To obtain more information about the 3LD-shaped h-BN, selected area electron diffraction (SAED) was used to analyze its crystal structure. Fig. 3a shows a low-magnification TEM image of a 3LD-shaped h-BN film. Four different positions, which are labeled in red in Fig. 3a, in the same dart were chosen, one from each leaf and one from the center. The SAED images of these positions are shown in Fig. 3b-e. These images clearly show the hexagonal structure of h-BN; in addition, the SAED images are rotationally aligned with each other, which suggests the single-crystalline nature of this 3LD-shaped h-BN domain.

In h-BN 2D crystals, there are three typical edges, which are N-terminated zigzag and B-terminated zigzag and armchair. N-terminated zigzag edges have been observed for most RT-shaped h-BN crystals synthesized by CVD on Cu substrates.[38-40] A dark-field image of an h-BN crystal is displayed in Fig. 4a, along with its corresponding diffraction pattern in the inset. The red line indicates the armchair direction of the h-BN crystal and the green line indicates the zigzag direction of the h-BN crystal, which were determined from the SAED image. To further determine the atomic arrangement (B or N position) of h-BN, aberration-corrected TEM at 80 kV was used to collect atomic images of the samples. Triangular holes with N-terminated zigzag edges, which could be used to evaluate the crystalline directions and monolayer configuration, were observed in the BN films,[41-43] as shown in Fig. 4b. Using the SAED patterns and atomic arrangement of B and N atoms, all the edge atom arrangements of the three non-RT shapes were determined; they are shown in Fig. 4c. The edges of the needle tail and V-shaped dart base possess N-terminated zigzag configuration, which is the same as that of RT h-BN. In contrast, the sides of the needle, V-shaped dart, and 3LD all have smooth or stepped armchair edges. To form a stable triangle, which is roughly the shape of the needles and dart leaves we found in the samples, when one side is a consecutive and smooth armchair, the other side has to be discrete armchairs separated by a number of vicinal steps; in other words, stepped armchair configuration. Considering the atom arrangement in a honeycomb structure, the angle between a smooth armchair edge and zigzag edge may be 30°, 90°, or 150°. In the SEM images, we did observe needles/dart leaves with one side edge (red lines in Fig.4d–f) perpendicular to the zigzag direction, which is perpendicular to one of the RT h-BN edges (each red line is perpendicular to the connecting green line in Fig. 4d–f). This finding is consistent with the above discussion about the edges of 3LD-shaped h-BN. Namely, the needles/dart leaves are right triangles.

Considering the different growth conditions using the two types of substrates, the oxygen content of the substrates attracted our attention. To explore the effect of substrate oxygen content on the formation of 3LD-shaped h-BN, we designed a series of experiments. LO-Cu substrates were exposed to oxygen (or double the time in air) by tuning the annealing time without $H_2$ from 0 to 30 min before adding any precursors. We found that triangular h-BN flakes formed when the exposure time was less than 10 min. Only after exposing the substrate to oxygen for longer than 10 min did 3LD-shaped h-BN form. As the oxygen exposure time lengthened, more 3LD-shaped h-BN flakes were found (Fig. 5a–d). Based on these results, oxygen is an important factor for the formation of 3LD-shaped h-BN. The role of oxygen in the growth process of 3LD-shaped h-BN will be discussed below.

The domain density is a crucial factor determining the quality of 2D films. The SEM images in Fig. 2a and b revealed that the h-BN domain densities on HO-Cu substrates were clearly lower than those on LO-Cu substrates under the same growth conditions. The calculated domain density of h-BN on LO-Cu was 0.044 μm$^2$, which was eight times larger than the 0.0054 μm$^2$ calculated for h-BN on HO-Cu. Based on growth theories, the defects on the metal surface are believed to serve as active sites for atomic nucleation in 2D material growth.[44] When growing graphene on a Cu substrate,[45,46] the high d-band centers at the low-coordinate defect sites lead to strong binding to adsorbates. For the same reason, O atoms in the ambient environment can also be trapped at these active sites, which slows the rate of nucleation. In the HO-Cu system, the active sites for h-BN nucleation can be occupied by O atoms, resulting in a lower domain density than that for the LO-Cu system. That is, the oxygen on the Cu surface suppressed h-BN nucleation in the HO-Cu system.

Besides the effects of suppressing h-BN nucleation, we also considered the influence of oxygen on the formation of the 3LD shape. Here, we compare the 3LD and triangle shapes based on their growth results. It is clear that both shapes have threefold rotation symmetry. However, the fast and slow growth directions in the LO-Cu system are $[0\,1\,\bar{1}\,0]$ and $[1\,0\,\bar{1}\,0]$, respectively, whereas the case is reversed in the HO-Cu system (Fig. 6a–c). In addition, there is another major difference between the HO-Cu and LO-Cu systems. The h-BN flakes formed on LO-Cu are compact triangles with sharp edges, which typically result from attachment kinetics.[47,48] For the h-BN flakes formed on HO-Cu, the growth domains are multi-branched or dendritic, which mainly arise from diffusion-controlled processes.[49-51] Therefore, the growth of h-BN changed from attachment kinetics to diffusion-controlled when the oxygen content of the substrate was increased.

During the growth of 2D materials, the variation of temperature and characteristic attachment time can substantially affect the growth kinetics.[33,52] The characteristic attachment time is affected by the edge energy barrier: the lower the edge energy barrier, the shorter the characteristic attachment time. Comparing the experimental results for different systems revealed that oxygen lowered the edge attachment barrier of h-BN and shortened the characteristic attachment time, which affected the growth kinetics and morphology of h-BN flakes. To investigate this behavior further, we performed density functional theory (DFT) calculations (See S4 for calculation detail) to analyze the edge barrier changes of h-BN.

Considering the H energy in hydrogen, the binding energy of hydrogen on a Cu surface is defined as:[52] $E(H@X) = E(H+X)-E(X)-E(H_2)/2$, where $X$ represents LO-Cu(111)/HO-Cu(111)/N-terminated h-BN, E(H+X) is the total energy of an H-saturated system that contains both H and X, E(X) is the energy of X without H, and $E(H_2)$ is the energy of a free $H_2$ molecule. During the CVD growth of h-BN, the borane-ammonia precursor was dehydrogenated to form BN (e.g., $BNH_x \xrightarrow{Cu} BNH_{x-1}+H$). In the HO-Cu system, oxygen on the Cu(111) surface can participate in the dehydrogenation process (e.g., $BNH_x+O \xrightarrow{Cu} BNH_{x-1}+OH$). Our DFT calculations, as shown in Fig. 6d, revealed that the binding energy of hydrogen on HO-Cu(111) was lower than that on LO-Cu(111) by 0.48 eV, implying a lower activation energy for dehydrogenation at h-BN edges based on the Bell–Evans–Polanyi principle.[53] At the edge of h-BN flakes, h-BN grows through dehydrogenation and continued attachment of B and N atoms. The edge attachment barrier decreased as the dehydrogenation barrier lowered. In other words, attachment kinetics was the main factor controlling growth in the LO-Cu system. Conversely, oxygen lowered the h-BN edge attachment barrier and promoted the growth process in the HO-Cu system. The decrease of the edge attachment barrier led to a short characteristic attachment time and changed the growth to a diffusion-controlled mode.

Closer examination of the SEM results (Fig. 6e) showed that there were a few triangular h-BN flakes with embossments on each side. Considering the above discussion, this is because at the beginning of h-BN growth, h-BN forms triangle-shaped flakes, but oxygen changed the h-BN growth kinetics in the HO-Cu system. As a result, the flakes grew faster along the direction perpendicular to the zigzag edges compared with the case in the LO-Cu system, producing sides embossed with humps that gradually became the leaves of 3LD-shaped h-BN (as shown in Fig. 6a to c). Based on the above discussion, the fast growth perpendicular to the zigzag edges resulted in the formation of 3LD-shaped h-BN. Furthermore, we can also explain the formation of needle- and V-shaped h-BN. In the HO-Cu system, the distribution of oxygen was not uniform. Some areas have relatively few oxygen atoms, which may be depleted gradually during growth and cause one or two missing leaves in the 3LD crystals. This oxygen depletion process led to the formation of needle-, V-, and triangle-shaped h-BN flakes. This behavior is also evidenced in the HO-Cu system with higher oxygen concentration, which contains less needle-, V-, and triangle-shaped flakes than the LO-Cu system.

**Conclusions**

We synthesized large single-crystal h-BN flakes with 3LD rather than triangular shape using an APCVD technique. Experimental analysis and DFT calculations revealed that oxygen plays an important role in the formation of 3LD-shaped h-BN. The quality of the h-BN was confirmed by XPS and EELS. Oxygen suppressed h-BN nucleation and decreased the h-BN domain density. SAED results confirmed that the 3LD-shaped h-BN was single crystalline. The HRTEM results showed that the edges of the 3LD-shaped h-BN are grouped armchairs, rather than the N-terminated zigzag edges of triangular h-BN. Oxygen lowered the attachment barrier at the edges of h-BN, which led to diffusion-controlled growth and formation of 3LD-shaped h-BN. The different shapes of h-BN with different edges produced here illustrate the role of oxygen in h-BN growth. The 3LD-shaped h-BN can serve as gate dielectrics for specially shaped field-effect transistors (FETs), especially ones based on 2D materials. Furthermore, the tip of 3LD-shaped h-BN, which is composed of only a few atoms, could be used to fabricate short-channel FETs. This study may open up a pathway to fabricate uniquely shaped and short-channel graphene and graphene-like FETs.

**Supplementary material**

See the supplementary material for the experimental setup (Fig. S1) and details, XPS O 1s analysis (Fig. S2), high-resolution TEM images (Fig. S3), and a dark–field image of needle-shaped h-BN crystals (Fig. S4).

**Conflicts of interest**

The authors declare no competing interests.

**Acknowledgments**

This work was supported by the National Key Research and Development Program of China (Grant No. 2016YFA0301700 and 2018YFA0306100), the National Natural Science Foundation of China (Grant No. 11625419, 51772265, and 51761165024), the National Basic Research Program of China (Grant No. 2014CB923500), the Anhui initiative in Quantum Information Technologies (Grant No. AHY080000). Part of this work was carried out at the USTC Center for Micro and Nanoscale Research and Fabrication and some TEM characterization was conducted at the Center of Electron Microscopy, Zhejiang University.

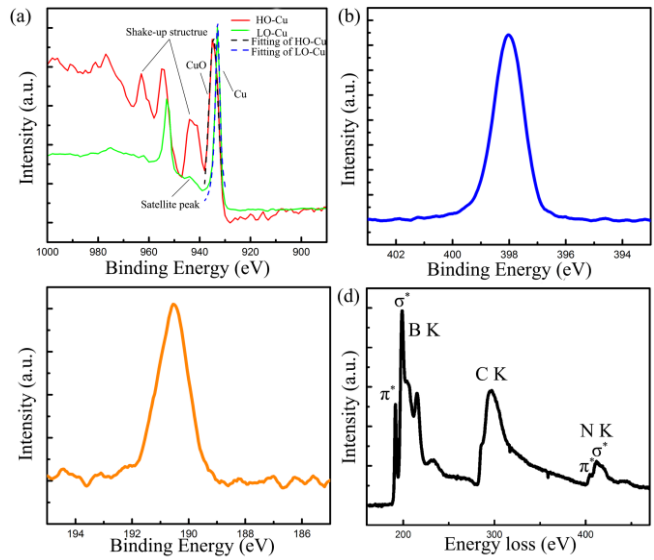

**Fig. 1** (a) O 1s spectra of HO-Cu and LO-Cu substrates before BN growth. (b) B 1s and (c) N 1s spectra of BN samples transferred onto 300-nm SiO$_2$/Si substrates. (d) Core EELS spectrum of a BN sample showing B and N K-shell excitation peaks. The K-shell excitation of C is from the ultrathin carbon film.

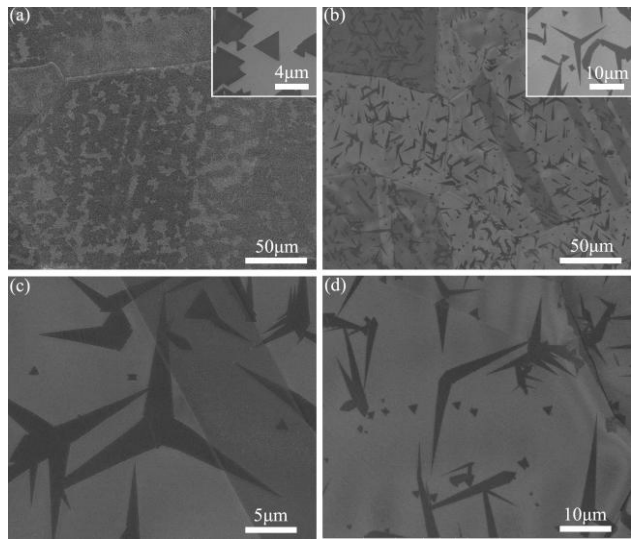

**Fig. 2** (a) SEM images of RT-shaped h-BN crystals grown on LO-Cu foil. (b) SEM images of 3LD-shaped h-BN crystals grown on HO-Cu foil. (c) High-resolution SEM images of 3LD-shaped h-BN. (d) High-resolution SEM images of needle- and V-shaped h-BN.

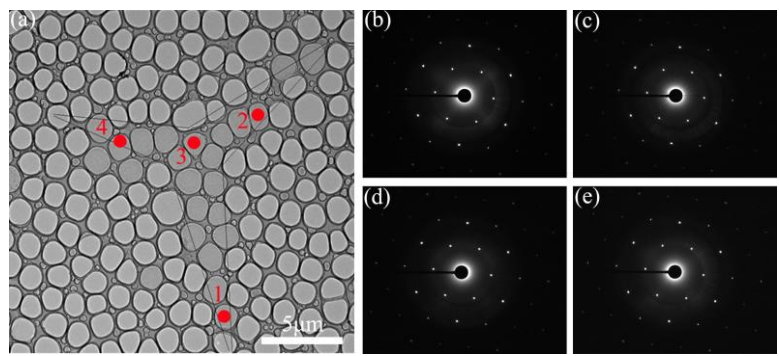

**Fig. 3** (a) TEM images of 3LD-shaped h-BN crystals grown on HO-Cu foil. (b to e) SAED images of the four positions labeled 1 to 4 in red in Fig. 3a.

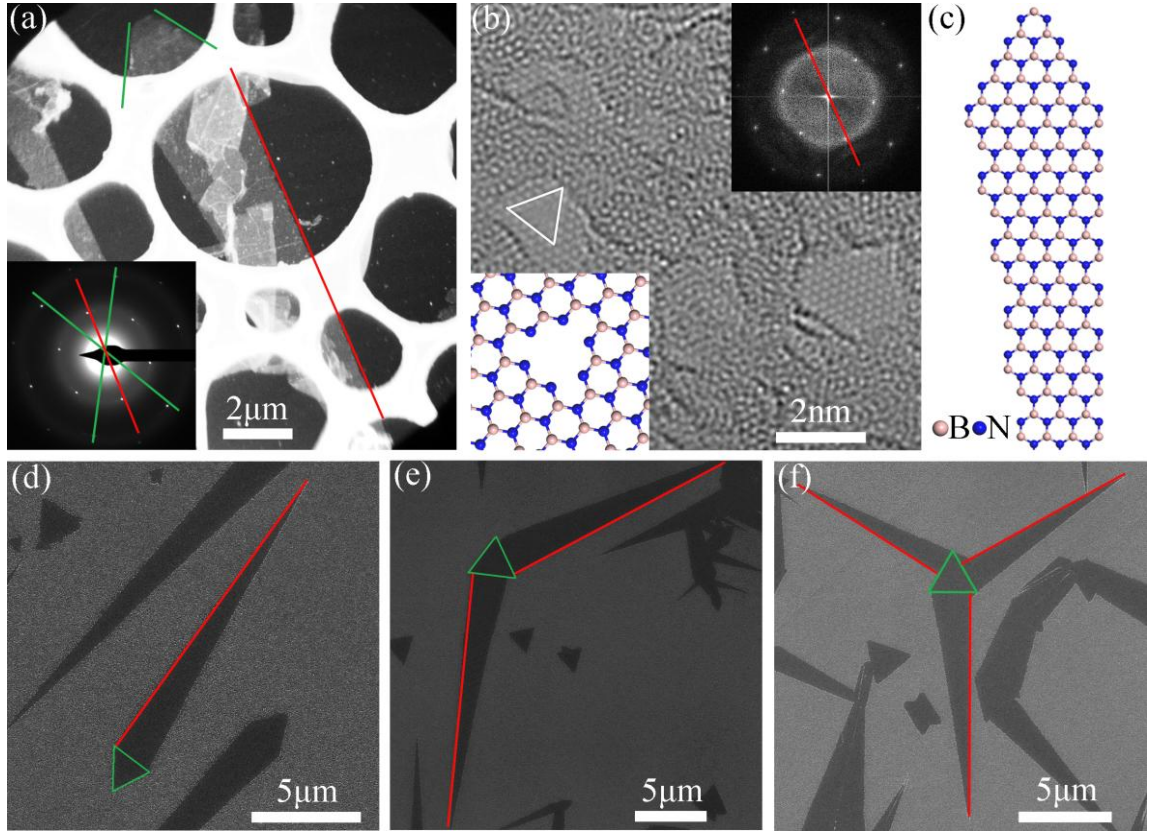

**Fig. 4** (a) Dark-field image of a needle-shaped h-BN crystal. Inset is the corresponding diffraction pattern. The red line indicates the armchair direction of the crystal. Note that there is no magnetic rotation between the image and diffraction pattern. (b) Corresponding atomic-resolution image of h-BN near the smooth edges, with triangular holes terminated by nitrogen atoms. The right inset is a fast Fourier transform of the HRTEM image. The red line indicates the armchair direction. The left inset is the atom arrangement model of the hole. (c) Schematics showing the atomic arrangement in 3LD-shaped h-BN. SEM images of (d) needle-, (e) V-, and (f) 3LD-shaped h-BN. Green triangles are RTs and red lines indicate edges with smooth armchair configuration.

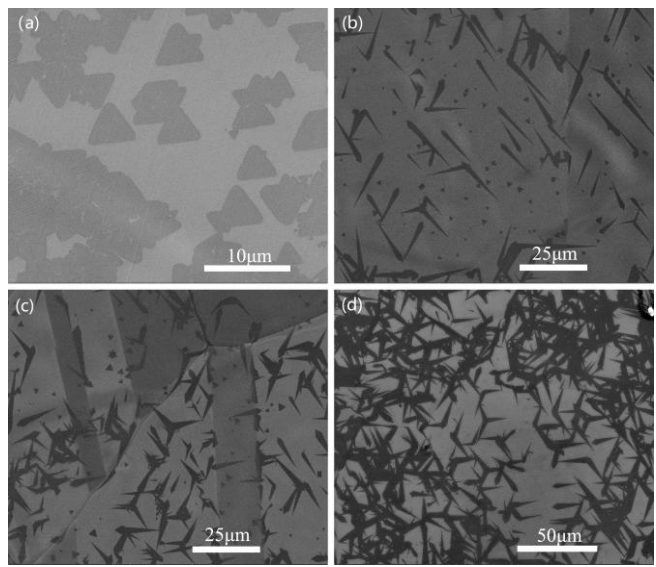

**Fig. 5** SEM images of h-BN crystals grown on LO-Cu exposed to oxygen for (a) less than 10 min, (b) 15 min, (c) 20 min, (d) 30 min.

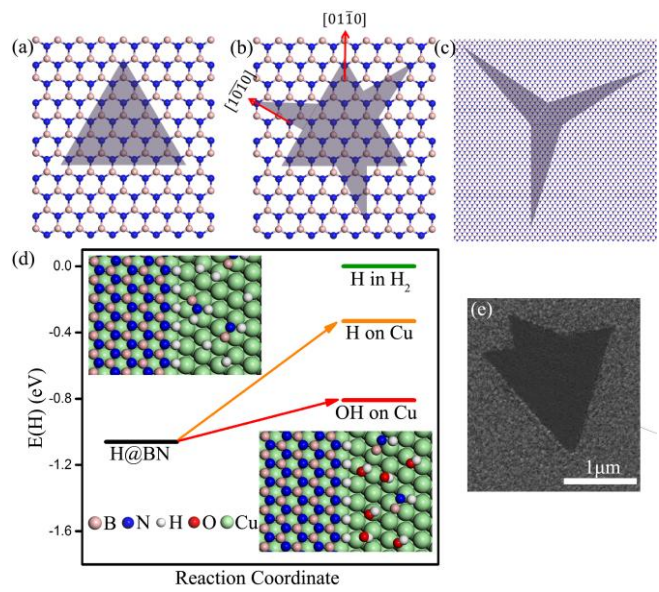

**Fig. 6** (a to c) Schematics of growth directions of h-BN domains. (d) DFT-calculated H-attachment binding energies for different configurations. The insets in top-left and bottom-right indicate h-BN edge growth on Cu with and without the assistance of O, respectively. (e) SEM image of triangular h-BN flake with embossments.